# Modeling and Experiments of an Injection-Locked Magnetron With Various Load Reflection Levels

Xiaojie Chen, *Graduate Student Member, IEEE*, Bo Yang, *Student Member, IEEE*, Naoki Shinohara, *Senior Member, IEEE*, and Changjun Liu, *Senior Member, IEEE*

***Abstract*— In this article, we investigate the performance of an injection-locked 5.8-GHz continuous-wave magnetron with various load reflection levels. The load reflection is introduced to an equivalent magnetron model to theoretically evaluate the system performance. The effects of different load reflection levels on the magnetron's output are numerically analyzed. Experiments are performed while the load reflection is varied using an E–H tuner between a magnetron and a circulator. A narrower locking bandwidth is observed under constant injection power with increasing load reflection. The proper-mismatched system suppresses its sideband energy, thereby reducing phase noise. The experimental features qualitatively validate the theoretical analyses results. The investigation results also provide guidance for advanced applications in communication and high-energy physics based on injection-locked magnetrons.**

## I. Introduction

MANY researchers have recently become interested in injection-locked magnetrons due to the advantages of low-cost, high-power, and small-volume, and developed advanced applications based on them. Shinohara [1] developed microwave phased arrays for wireless power transmission based on phase-controlled, injection-locked magnetrons. Dexter *et al.* [2] demonstrated that a superconductor cavity of a particle accelerator is precisely driven by an injection-locked magnetron. Tahir *et al.* [3] and Yang *et al.* [4] investigated the communication applications of the modulated injection-locked magnetrons. Audio and video signals were wirelessly transmitted and well displayed in Yang's demonstration [4]. Furthermore, Liu *et al.* [5], [6] concentrated on a high-power and -efficiency microwave power combining system based on multiway injection-locked industrial magnetrons to meet the increasing power requirement of industrial applications. Therefore, the magnetron performance must be estimated before usage in practical applications.

The Reike diagram is currently being widely adopted to portray the operating status of magnetrons; however, it cannot demonstrate the variations of the locking bandwidth and output quality. Adler's condition is commonly used to predict the locking bandwidth [7]; the formula is as follows:

$$\Delta f \leq \rho f_0/(2Q_{\mathrm{ext}}) \tag{1}$$

where $\rho$, $\Delta f$, $f_0$, and $Q_{\mathrm{ext}}$ are the injection ratio, locking bandwidth, resonant frequency, and external quality factor, respectively. The locking bandwidth and the output quality are susceptible to variations in load impedance [8], [9]. However, the tendency of the locking bandwidth and output quality variations has not yet been clearly explained.

This study symbolizes load variation by a reflection coefficient and proposes a model to illustrate the performance of an injection-locked magnetron based on the equivalent circuit and scattering network. We developed a 5.8-GHz injection-locked magnetron system with an E–H tuner between a magnetron and a circulator. The locking bandwidth and phase noise are measured with various load reflection levels introduced by the E–H tuner. The experimental results qualitatively agree with the theoretical estimations. Herein, suitable strategies are presented considering various applications based on injection-locked magnetrons.

## II. Theory of the Injection-Locked Magnetron With Various Load Reflection Levels

### A. Locking Bandwidth With Various Load Reflection Levels

A $\pi$-mode operating magnetron can be equivalent to a parallel resonant circuit comprising lumped components (see Fig. 1). The energy is generated through electron admittance represented as $g_e + jb_e$. The magnetron resonance cavity

Manuscript received June 12, 2020; accepted July 12, 2020. Date of publication August 3, 2020; date of current version August 21, 2020. The work of Xiaojie Chen was supported in part by the China Scholarship Council under Grant 201906240240, in part by NSFC under Grant 61931009, and in part by the China 973 Program under Grant 2013CB328902. The review of this article was arranged by Editor M. Blank. *(Corresponding authors: Naoki Shinohara; Changjun Liu.)*

Xiaojie Chen is with the School of Electronics and Information Engineering, Sichuan University, Chengdu 610064, China, and also with the Research Institute for Sustainable Humanosphere, Kyoto University, Uji 611-0011, Japan (e-mail: xjchen9112@163.com).

Bo Yang and Naoki Shinohara are with the Research Institute for Sustainable Humanosphere, Kyoto University, Uji 611-0011, Japan (e-mail: yang_bo@rish.kyoto-u.ac.jp; shino@rish.kyoto-u.ac.jp).

Changjun Liu is with the School of Electronics and Information Engineering, Sichuan University, Chengdu 610064, China (e-mail: cjliu@ieee.org).

Color versions of one or more of the figures in this article are available online at http://ieeexplore.ieee.org.

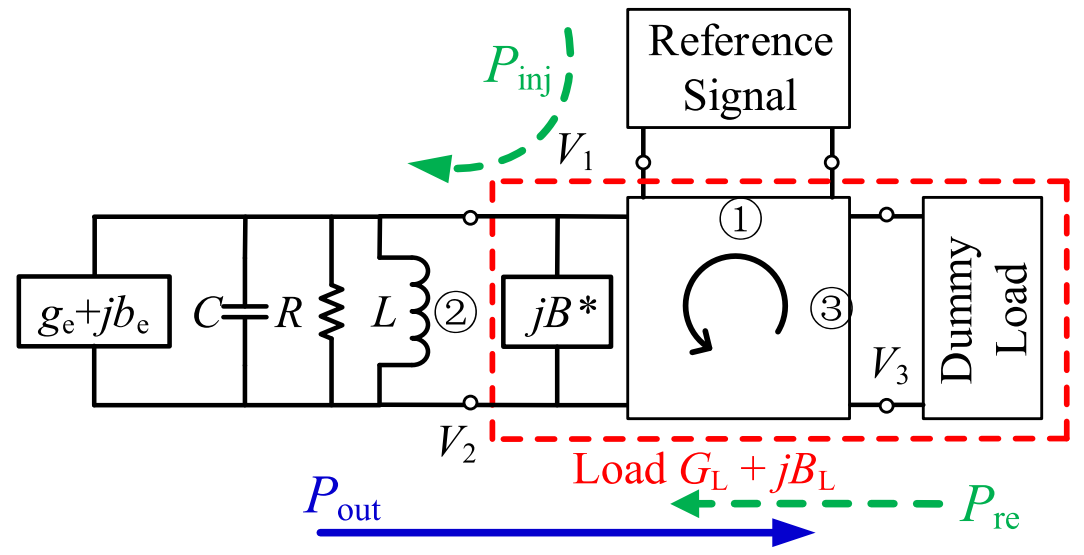

Fig. 1. Schematic of the injection-locked magnetron system with variable load reflection.

is approximated using a resistance–inductance–capacitance ($RLC$) shunt resonant circuit. The load $G_{\mathrm{L}} + jB_{\mathrm{L}}$ is formed using an susceptance tuner, a circulator, and a dummy load. The susceptance tuner $jB^*$ between the magnetron and the circulator is supposed to introduce various susceptances. The equilibrium oscillation requirement is expressed as follows:

$$\begin{cases} g_e = -1/R - G_{\mathrm{L}} \\ jb_e = -j\left(\omega_0 C - 1/(\omega_0 L)\right) - jB_{\mathrm{L}}. \end{cases} \tag{2}$$

The output RF voltage and external quality factor of a magnetron system, respectively, are given as follows [10], [11]:

$$V_{\mathrm{RF0}} = V_{\mathrm{dc}}/\left(\omega_0 RC\left(1/Q_{\mathrm{L}} + 1/RC\omega_0\right)\right) \tag{3}$$
$$Q_{\mathrm{ext}} = \omega_0 C/G_{\mathrm{L}} \tag{4}$$

where $V_{\mathrm{dc}}$ is the anode voltage; $Q_{\mathrm{L}}$ is the load quality factor; $b_0$ is a constant. Substituting (2), $Q_{\mathrm{L}} = \omega_0 C/g_e$, and $Q_0 = \omega_0 CR$ into (3), the RF voltage of the magnetron can be rewritten as

$$V_{\mathrm{RF0}} = V_{\mathrm{dc}}/(R \cdot g_e + 1) = V_{\mathrm{dc}}/\left(R\left(-G_{\mathrm{L}} - 1/R\right) + 1\right). \tag{5}$$

We assumed that a magnetron was initially working at 5.8 GHz with a perfect match load. Its $Q_{\mathrm{ext}}$, $Q_0$, $G_{\mathrm{L}}$, and $B_{\mathrm{L}}$ were equal to 50, 1200, 1 S, and 0 S, respectively. We consider that a varied load susceptance $B'_{\mathrm{L}}$ causes a load reflection $|\Gamma|$. The $B'_{\mathrm{L}}$ can be determined using $Y'_{\mathrm{L}} = Y_0(1 - |\Gamma|)/(1 + |\Gamma|)$

$$1 + jB'_{\mathrm{L}} = (1 - |\Gamma|)/(1 + |\Gamma|) \rightarrow jB'_{\mathrm{L}} = -2|\Gamma|/(1 + |\Gamma|). \tag{6}$$

Thereby, the renewed $Y'_{\mathrm{L}}$ is real and considered as the renewed conductance $G'_{\mathrm{L}}$

$$Y'_{\mathrm{L}} = G_{\mathrm{L}} + jB' = 1 - 2|\Gamma|/(1 + |\Gamma|) = G'_{\mathrm{L}}. \tag{7}$$

The renewed RF voltage and external quality factor can be obtained by substituting (7) into (3) and (4), respectively. We then obtain the following equations:

$$V'_{\mathrm{RF0}} = V_{\mathrm{dc}}/\left(R\left(-G'_{\mathrm{L}} - 1/R\right) + 1\right) \tag{8}$$
$$Q'_{\mathrm{ext}} = \omega_0 C/G'_{\mathrm{L}}. \tag{9}$$

Then the renewed operating conditions of the magnetron (with nonperfect matched load) can be expressed as

$$V'_{\mathrm{RF0}} = V_{\mathrm{RF0}}/k \tag{10}$$
$$Q'_{\mathrm{ext}} = Q_{\mathrm{ext}}/k \tag{11}$$

where $k = 1 - 2|\Gamma|/(1 + |\Gamma|)$.

A widely applied schematic [2]–[6] of the injection-locking magnetron system is shown in Fig. 1, where a circulator connected to the magnetron to provide the injected path for the reference signal. Herein, the scatting matrix for the injection-locked magnetron with a mismatched three-port circulator (see Fig. 1) is given by [12]

$$\begin{bmatrix} V_1^- \\ V_2^- \\ V_3^- \end{bmatrix} = \begin{bmatrix} |\Gamma| & |\Gamma| & \alpha \\ \alpha & |\Gamma| & |\Gamma| \\ |\Gamma| & \alpha & |\Gamma| \end{bmatrix} \begin{bmatrix} V_1^+ \exp\left(-j\omega_{\mathrm{inj}}t - j\chi_1\right) \\ V_2^+ \exp\left(-j\omega_{\mathrm{inj}}t - j\chi_2\right) \\ V_3^+ \exp\left(-j\omega_{\mathrm{inj}}t - j\chi_3\right) \end{bmatrix} \tag{12}$$

where $V_i^-$ and $V_i^+$ are the output waves and input waves, respectively; $\omega_{\mathrm{inj}}$ is the angular frequency of the injected signal; $\chi_i$ is the port phase of the circulator; and $\alpha = 1 - |\Gamma|^2$. $S_{22}$ in (12) changes the operating condition of the magnetron, as aforementioned in (10) and (11). In Fig. 1, the blue solid arrow indicates the magnetron's output power streams to the load, whereas the green dash arrows indicate the power streams into the magnetron. The injected RF voltage to the magnetron $V_2^+$ can be written as

$$V_2^- = S_{21}V_1^+ e^{\left(-j\omega_{\mathrm{inj}}t - j\chi_1\right)} + S_{23}V_3^+ e^{\left(-j\omega_{\mathrm{inj}}t - j\chi_3\right)} \tag{13}$$

where $V_3^+ = S_{32}S_{33}V_2^+$. Then the injection ratio can be written as

$$\boldsymbol{\rho}' = \left(V_1^+ S_{21} e^{j(\omega_{\mathrm{inj}}t + \chi_1)} + V'_{\mathrm{RF0}} S_{32} S_{33} S_{23} e^{j(\omega_{\mathrm{inj}}t + \chi_3)}\right) / \left(V'_{\mathrm{RF0}} e^{j\omega t}\right). \tag{14}$$

In (14), when locking occurs, the phase difference between $\chi_1$ and $\chi_3$ is a constant value. Substituting (8) into (14), we then obtain

$$\boldsymbol{\rho}' = \rho k S_{21} e^{j(\Delta\omega t + \chi_1)} + \beta S_{32} S_{33} S_{23} e^{j\Delta\omega t} \tag{15}$$

where $\rho$ is the initial injection ratio of the perfect matched condition, and $\Delta\omega$ is $\omega_{\mathrm{inj}} - \omega$. The phase of $\rho'$ is unity to $\chi_1$; thus, $\beta = \cos\chi_3$ is the projection parameter whose value is within the limitation of $\beta \in [0, 1]$. The effective amplitude of the injection ratio is presented as

$$|\boldsymbol{\rho}'| = \alpha\sqrt{\rho^2 k^2 + \beta^2 |\Gamma|^4 + 2\beta\rho k |\Gamma|^2 \cos\chi_1}. \tag{16}$$

The initial phase of the magnetron output is assumed to be zero, such that $\chi_1$ is the phase difference between $\boldsymbol{\rho}'$ and the magnetron's free-running output. Thus, the locking bandwidth formula is rewritten as

$$\Delta f' = |\boldsymbol{\rho}'| f_0 \sin(\chi_1) / \left(2Q'_{\mathrm{ext}}\right). \tag{17}$$

However, (17) has become a transcendental equation. Hence, we assumed the value-limited item (i.e., $\cos\chi_1$) is represented by a variable $\gamma$, whose value is within $\gamma \in [-1, 1]$. Accordingly, (17) can be written as

$$f(\gamma) = \gamma^3 + \varepsilon\gamma^2 - \gamma - \varepsilon + \mu = 0 \tag{18}$$

where $\varepsilon = k\rho/(2\beta|\Gamma|^2) + \beta|\Gamma|^2/(2k\rho)$, $\mu = 2\Delta f'^2 Q_{\mathrm{ext}}^2/(k^3\beta\rho f_0^2 \alpha^2 |\Gamma|^2)$, and $f(\pm 1) = \mu > 0$.

Next, we derived the following equation from (18):

$$f'(\gamma) = 3\gamma^2 + 2\varepsilon\gamma - 1 = 0. \tag{19}$$

The solutions for (19) (i.e., $\gamma_1$ and $\gamma_2$) indicate the corresponding extreme values of (19) obtained using the following equation:

$$\gamma_1 = -\varepsilon/3 + \sqrt{\varepsilon^2+3}/3,\ \gamma_2 = -\varepsilon/3 - \sqrt{\varepsilon^2+3}/3. \quad (20)$$

The $\varepsilon$ values are surely greater than 1 when substituting the preset data. The $\gamma_2$ value is certainly below $-1$, exceeding the interval of the $\gamma \in [-1, 1]$. However, $\gamma_1$ is exactly within the value range. The extreme value $f'(\gamma_1)$ should be unique and satisfy the limitation of $f'(\gamma_1) \le 0$ to guarantee that the solution exists in (19). Utilizing $\gamma_1$, (18) can be transformed into

$$\mu \le 2\varepsilon/3 - 2\varepsilon^3/27 + 2\left(\varepsilon^2+3\right)^{\frac{3}{2}}/27. \quad (21)$$

Finally, we obtain a renewed locking bandwidth formula

$$\Delta f' \le f_0\alpha|\Gamma|\sqrt{k^3\beta\rho}\sqrt{2\varepsilon/3-2\varepsilon^3/27+2\left(\varepsilon^2+3\right)^{\frac{3}{2}}/27}/\left(\sqrt{2}Q_{\text{ext}}\right). \quad (22)$$

We chose $\beta = 1/(2^{0.5})$ for the following numerical analysis, which is the root-mean-square value of $\cos\chi_3$.

### B. Phase Noise With Various Load Reflection Levels

The output microwaves of free-running magnetrons usually span a certain frequency band $\Delta f_b$ which can be estimated using $\Delta f_b = f_c/Q_L$ [13], where $f_c$ is the central frequency of the output spectrum, $Q_L$ is the loaded quality factor determined using $\eta_c = Q_L/Q_{\text{ext}}$ [14]. Considering the magnetron connects with a nonperfect matched load, then the renewed spectrum bandwidth is obtained using

$$\Delta f'_b = f_c/\left(\eta'_c Q'_{\text{ext}}\right) \quad (23)$$

where $\eta'_c = G'_L/(G'_L + Q'_{\text{ext}}/Q_0)$ [10] and the deviation of central frequency $f_c$ caused by the load-pull effect is omitted due to $\Delta f_c \ll f_c$.

Then the normalized unilateral spectrum bandwidth is determined using initial unilateral spectrum bandwidth as the reference

$$f_n = \frac{\Delta f'_b}{\Delta f_b} = \frac{k^2Q_0 + Q_{\text{ext}}}{k\left(Q_0+Q_{\text{ext}}\right)}. \quad (24)$$

The phase noise of the magnetron follows the ideal $1/f^2$ dependence:

$$|\tilde\delta_0(f_m)|^2 = \frac{10^{k_0}}{f_m^2} \quad (25)$$

where $f_m$ is the offset angular frequency migrated from $f_c$, $k_0$ is the index. The phase noise of free-running magnetron with varied load reflection coefficient can be read as

$$|\tilde\delta'_0(f_m)|^2 = \frac{10^{k'_0}}{f_m^2} = \frac{10^{\lg\left(|\tilde\delta_0(f_m)|^2 f_n^2 f_m^2\right)}}{f_m^2}. \quad (26)$$

When an external injection is introduced and the reference frequency equals to $f_c$ (the locked phase difference between reference signal and magnetron's output equals to $2n\pi$, where $n$ is an integer), the spectral density of the magnetron with respect to various reflection levels can be obtained [15]

$$|\tilde\delta\theta(\omega_m)|^2 = |\tilde\delta\theta'_0(\omega_m)|^2\frac{(\omega_m/\omega_{3\,\text{dB}})^2}{|\boldsymbol{\rho}'|^2+(\omega_m/\omega_{3\,\text{dB}})^2} + |\tilde\delta\theta_{\text{inj}}(\omega_m)|^2\frac{|\boldsymbol{\rho}'|^2}{|\boldsymbol{\rho}'|^2+(\omega_m/\omega_{3\,\text{dB}})^2} \quad (27)$$

where $|\tilde\delta\theta_{\text{inj}}(\omega_m)|^2$ describes the spectral density of the external injection signal, $\omega_{3\,\text{dB}}$ equals to $\omega_0/(2Q'_{\text{ext}})$.

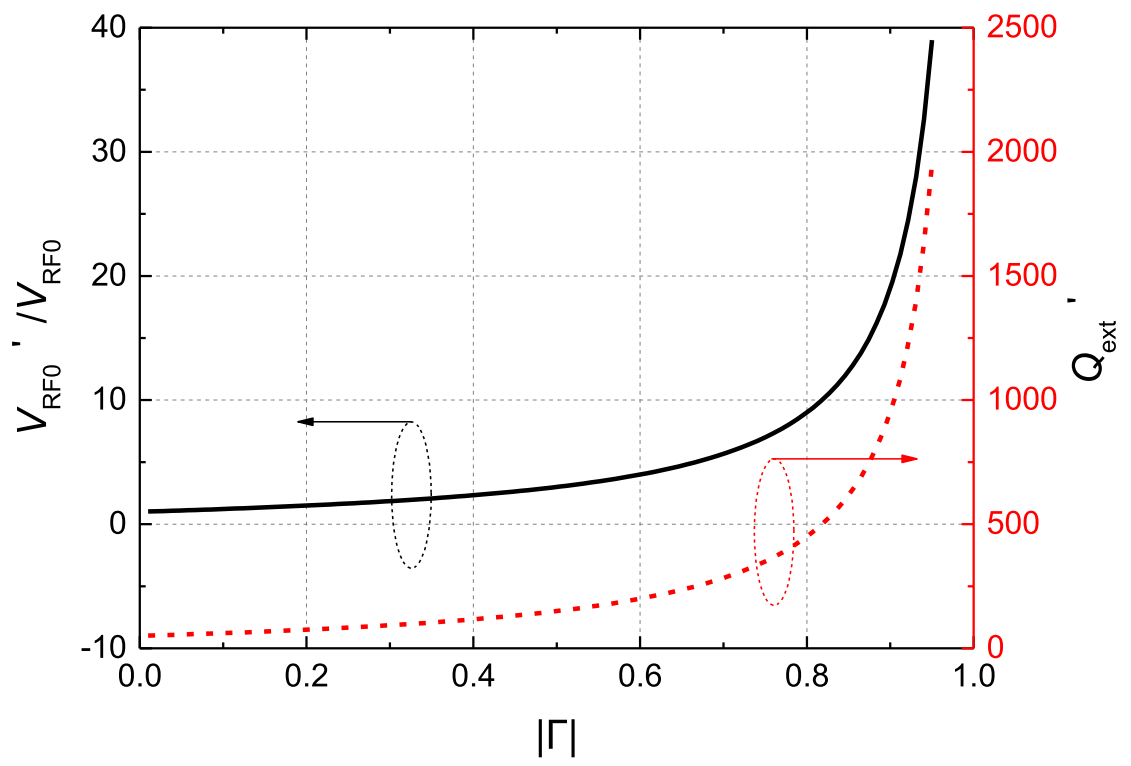

Fig. 2. Renewed output voltage and external quality factor versus varied load reflection level.

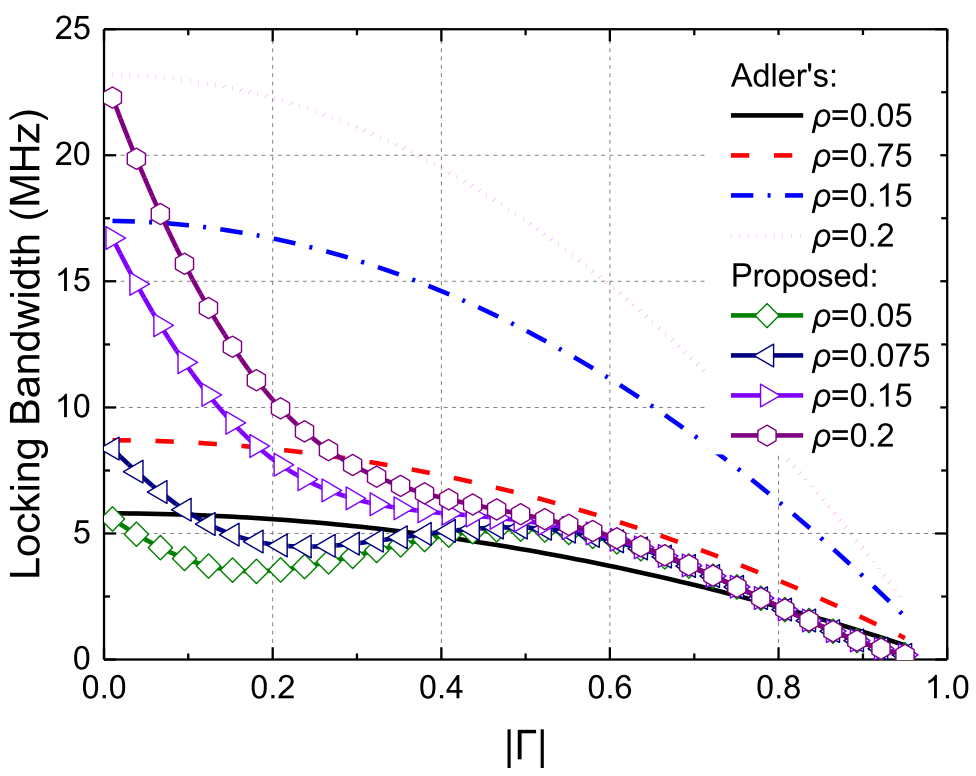

Fig. 3. Locking bandwidth characteristics of a magnetron with various load reflection levels.

### C. Numerical Calculation

We used the deduced equations in Sections II-A and II-B in the numerical computation to theoretically evaluate the performance of the injection-locked magnetron with various load reflection levels.

Variable $|\Gamma|$ was within the limitation of $|\Gamma| \in [0.01, 0.95]$. Fig. 2 depicts the variation tendency of the RF voltage ratio $V'_{\text{RF0}}/V_{\text{RF0}}$ and the external quality factor $Q'_{\text{ext}}$ with respect to various reflection levels using (10) and (11). $V'_{\text{RF0}}/V_{\text{RF0}}$ and $Q'_{\text{ext}}$ drastically increased when $|\Gamma|$ was close to 1. The $Q'_{\text{ext}}$ curve indicates that the coupled RF energy was drastically weakened. The magnetron no longer effectively outputs RF energy then the dc-to-microwave conversion efficiency will significantly decrease.

Accordingly, (22) suggests that the operating status of an injection-locked magnetron is actually altered by the variation of $V'_{\text{RF0}}$ and $Q'_{\text{ext}}$. The curves in Fig. 3 shows the locking bandwidth estimated by the proposed model and Adler's equation at similar instances. In the calculation of Adler's equation, the external quality factor $Q_{\text{ext}}$ is fixed: $Q_{\text{ext}} = 50$, but the injection ratio was varied with respect to the reflection level, i.e., $\rho(1 - |\Gamma|^2)$. Using the minimum reflection $|\Gamma| = 0.01$ in our calculation, the computed locking bandwidth was approximate to the prediction of Adler's equation. Compared to Adler's prediction, the locking bandwidth was sensitive to the reflection level. Set the locking bandwidth at $|\Gamma| = 0.01$ as the reference and using $(\Delta f_{\text{optimal}} - \Delta f_{|\Gamma|})\ /\ \Delta f_{\text{optimal}} \times 100\%$, the locking bandwidth substantially decreased to 26.2% ($\rho = 0.05$), 37.0% ($\rho = 0.075$), 59.0% ($\rho = 0.15$), and 75.2% ($\rho = 0.2$), respectively, when $|\Gamma|$ adjusted from 0.01 to 0.4. But the deviation of Alder's condition was fixed: 15.4%. With the reflection coefficient increasing further, the locking bandwidth continuously decreased but the slope of curves was becoming smaller than the condition of $|\Gamma| < 0.4$.

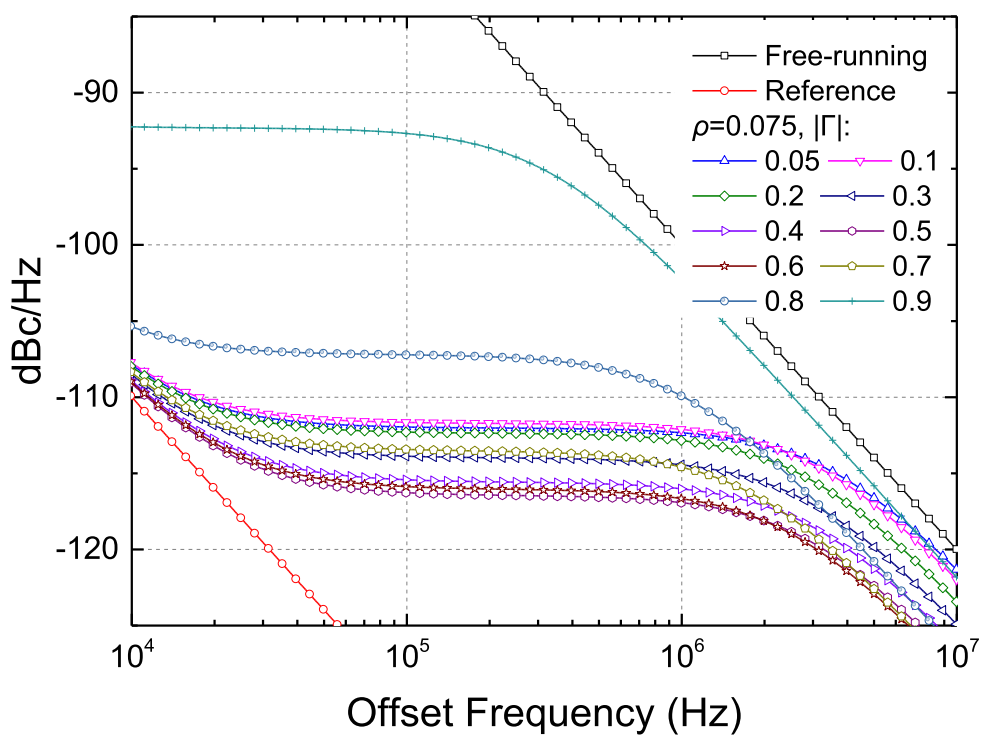

(a)

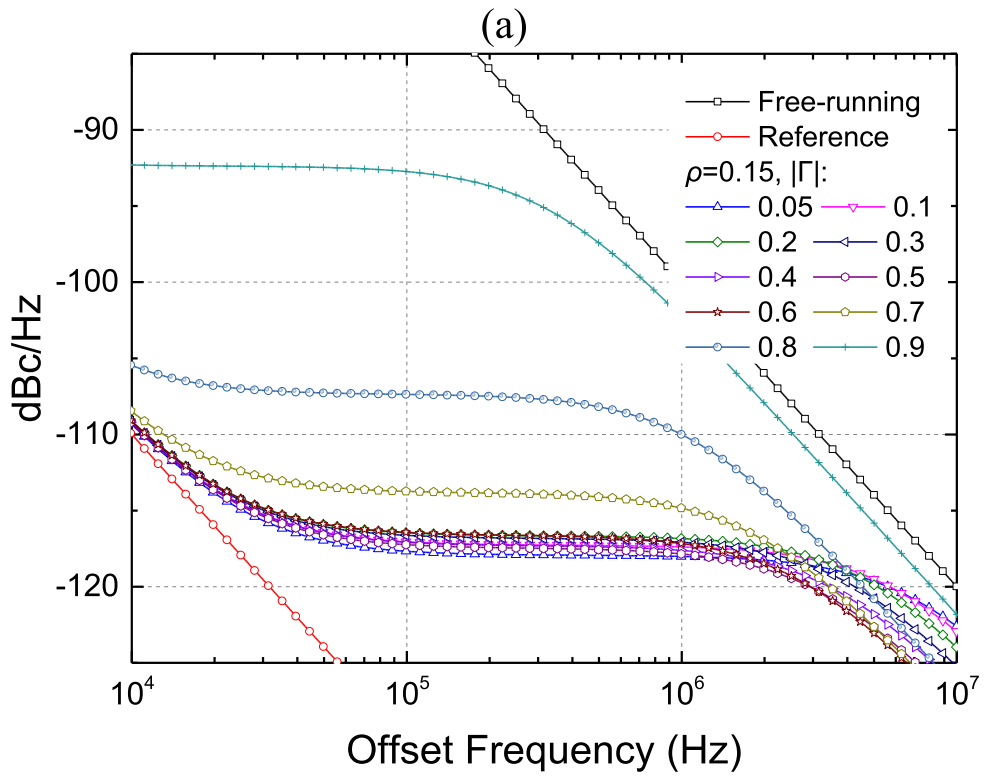

(b)

Fig. 4. Phase noise characteristics of the injection-locking magnetron with different load reflection levels. (a) $\rho = 0.075$. (b) $\rho = 0.15$.

In Fig. 4, the phase noise of the magnetron was cooperatively effected by the external injection and load reflection, as theoretically illustrated using (27). Both initial phase noises followed the ideal $1/f^2$ dependence: $|\tilde{\delta\theta}_0(f_m)|^2 = 10^2/(f_m)^2$ and $|\tilde{\delta\theta}_{\text{inj}}(f_m)|^2 = 10^{-3}/(f_m)^2$. When locking occurred, the phase noise level was suppressed. As shown in Fig. 4(a), with the reflection coefficient adjusting from 0.05 to 0.5, the noise level of injection-locked magnetron was reduced with a suppression of ~6 dB at the low offset frequency band. With the reflection coefficient approaching 0.7 and increasing even higher, the injection-locked magnetron is deteriorated with a level of ~20 dB. With a higher initial injection ratio of $\rho = 0.15$ [see Fig. 4(b)], noise levels was further suppressed, but the level difference of the phase noise curves with $|\Gamma| < 0.6$ is not so distinct. Furthermore, the phase noise levels at the higher offset frequency band (approaching 1 MHz and increasing even further) were also suppressed.

## III. Experimental System Setup

Additionally, we developed a corresponding experimental system to verify our theoretical analysis. Fig. 5 shows the block diagram and photograph of the system. The magnetron (model: M5802-KRSC1) was manufactured by Panasonic Microwave Co. (Japan) with a 5.8-GHz continuous wave output. The magnetron was driven using a switch-mode dc power supply (WepeX 1000B-TX, Megmeet) with an improved anode voltage ripple of ~1%. The filament current can be turned off after 5-min preheating. A relatively sharp free-running spectrum was achieved.

An oscilloscope (TDS-3054, Tektronix) was used to measure the anode voltage (high-voltage probe: P6015A, Tektronix) and current (ac/dc current probe: 1146A, Hewlett Packard). A reference signal was generated using a signal generator (N5172B, Keysight) and amplified using a power amplifier (CA5800BW50-4040R, R&K). The circulators provided a transmission path for the injection of the amplified reference signal, thereby protecting the solid-state amplifier. Couplers were used to sample the signals and measure the power and spectrum using power meters (A1914A, Agilent) and a signal analyzer (N9010A, Agilent), respectively. The output microwave power was absorbed by a dummy load.

Various load reflection levels were introduced through an E–H tuner (EMH-6H, Nihon Koshuha) placed between the magnetron and the circulator #1. The E–H tuner integrated the removable short-end plug at both $E$- and $H$-planes. Simultaneously, the load characteristic variation was measured using a vector network analyzer (N9928A, Keysight).

## IV. Experimental Results and Discussion

### A. Different Load Reflection Levels

The load characteristics were measured before connecting with the magnetron, as shown in Fig. 5(a). The red dashed line in Fig. 5(b) depicts the load part. Fig. 6 shows that various load performances were recorded by shifting the $H$-plane short-end plug of the tuner, while the offset depth of the $E$-plane short-end plug remained fixed. Nine sets of $|S_{11}|$ data with 3-dB intervals were recorded and using to deduced reflection coefficient $|\Gamma|$, the corresponding depths of the short-end plug in $H$-plane were also presented in Fig. 6; $|\Gamma|$ at 5.8 GHz varied from 0.06 to 0.98. The depth of the short-end plug and the varied $|\Gamma|$ at 5.8 GHz was recorded for later use and convenience.

### B. Performances Versus Load Reflection Levels

The magnetron operating properties were first measured with an optimal load reflection level (i.e., $|\Gamma| = 0.06$ in Fig. 6) and determined as the initial status. Two initial statuses (e.g., $P_{\text{out}} = 370$ and 180 W) were investigated.

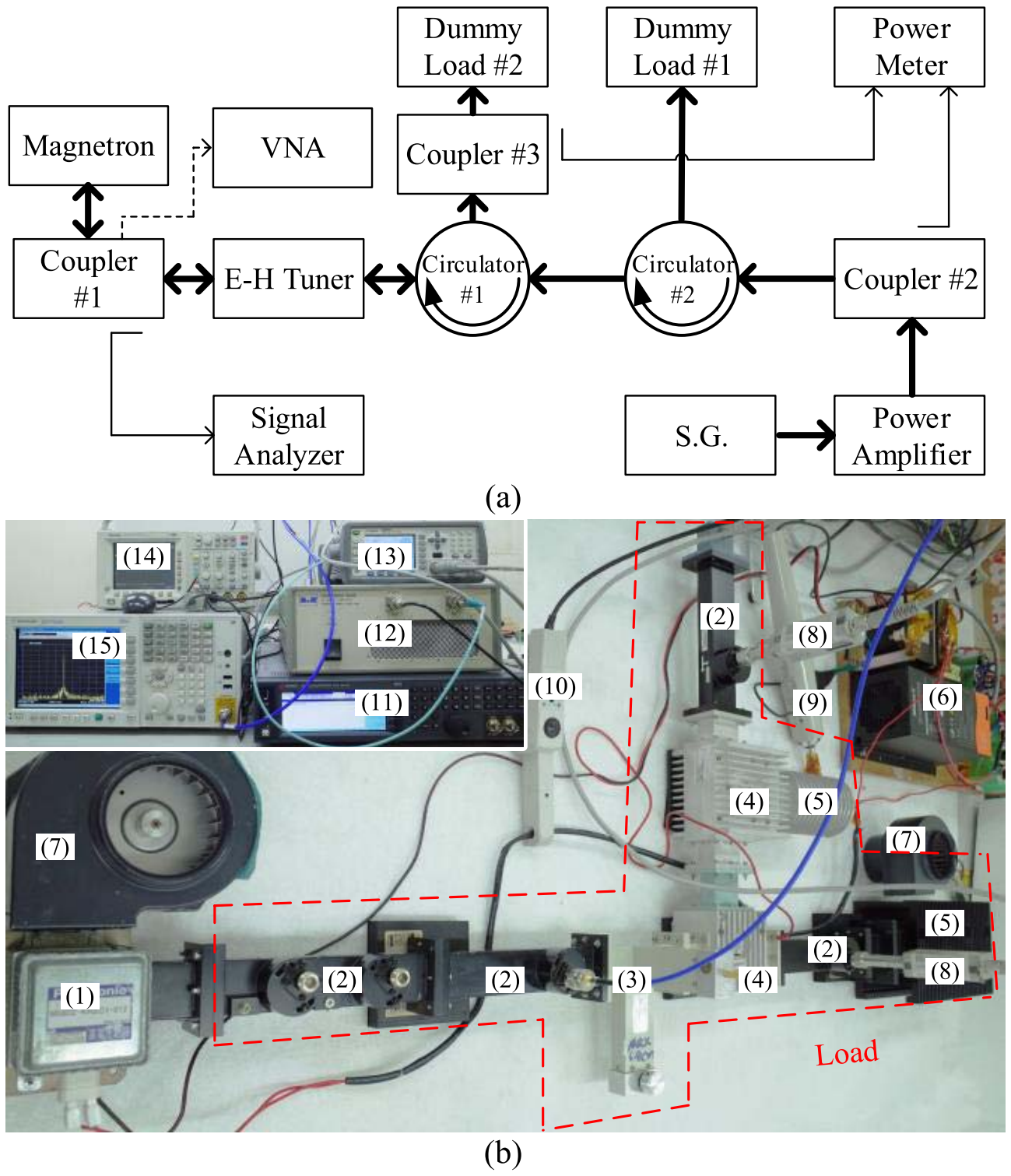

Fig. 5. (a) Block diagram and (b) photograph of the experimental system. Components and devices: (1) magnetron; (2) coupler; (3) E–H tuner; (4) circulator; (5) dummy load; (6) power supply; (7) fan; (8) power sensor; (9) high-voltage probe; (10) current probe; (11) signal generator; (12) power amplifier; (13) power meter; (14) oscilloscope; and (15) signal analyzer.

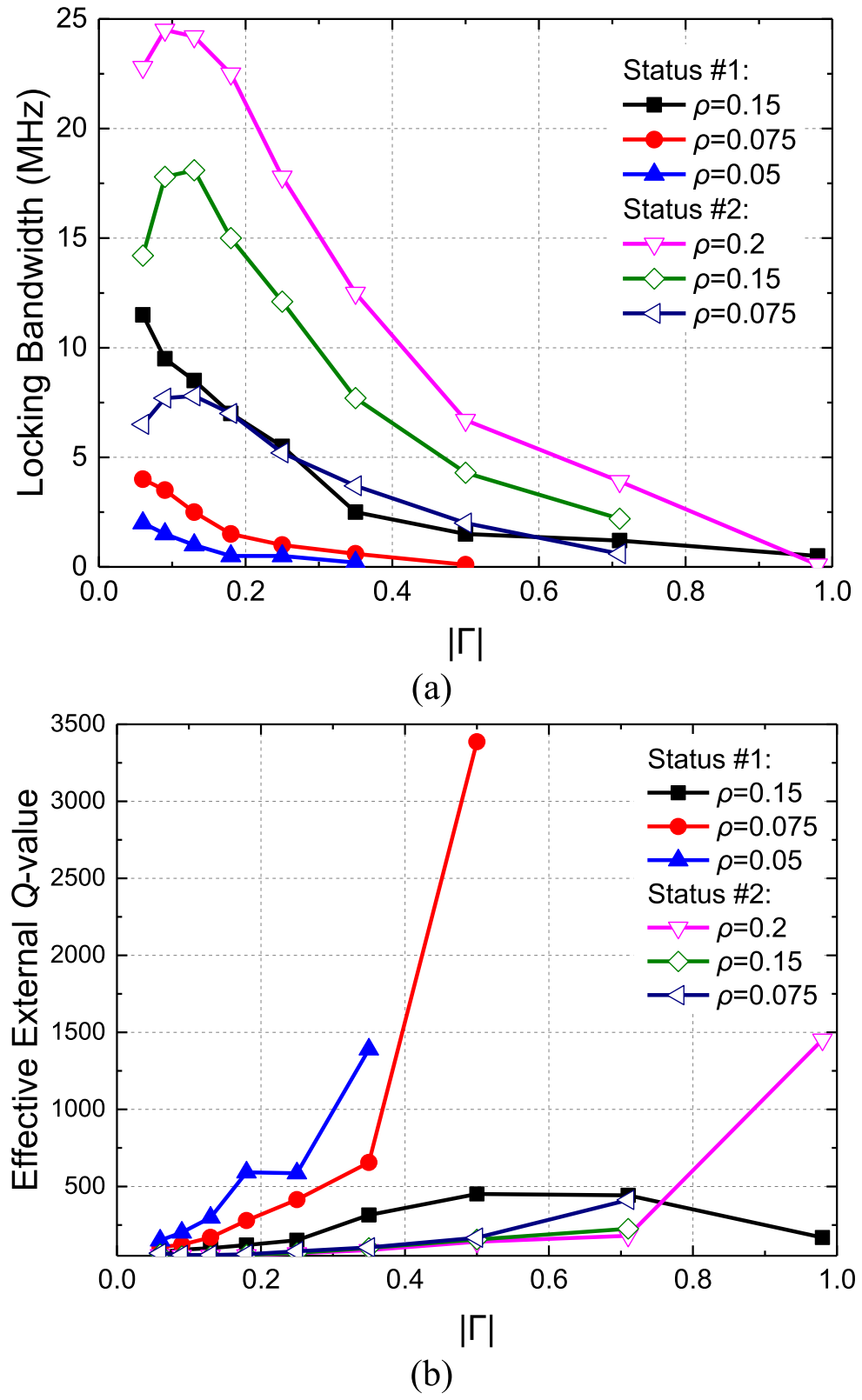

Fig. 7. (a) Measured locking bandwidth and (b) effective $Q_{ext}$ of the different load reflection levels.

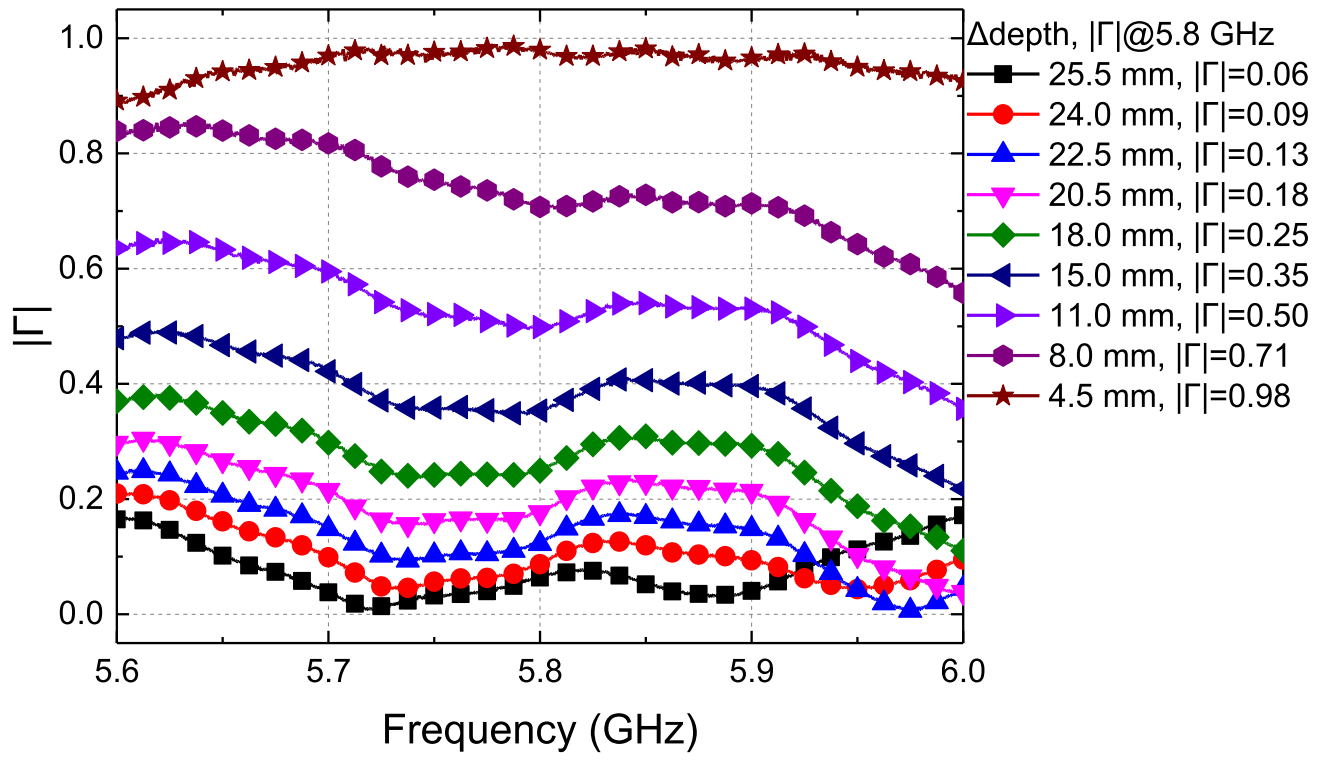

Fig. 6. Measured load reflection coefficient $|\Gamma|$.

Fig. 7(a) shows the evaluated locking bandwidth of the two statuses with various load reflection levels and the initial injection ratio $\rho = (P_{\text{in}}/P_{\text{initial}})^{1/2}$, where $P_{\text{in}}$ and $P_{\text{initial}}$ are the injected power and the output power of magnetron with the optimal load condition, respectively. In Status #1, when the load reflection effects were introduced, the locking bandwidth decreased drastically when $|\Gamma|$ varied from 0.06 to 0.35, the decrements were 90.0% ($\rho = 0.05$), 85.0% ($\rho = 0.075$), and 78.3% ($\rho = 0.15$) in Status #1, respectively. Contrarily, the locking bandwidth altered softly with $|\Gamma|$ further increasing.

Unlike Status #1, the measured results shown that the optimal load condition of Status #2 appeared when $|\Gamma| = 0.13$. This phenomenon was caused by the frequency pushing effect and the assembly error of the waveguide connection. The locking bandwidth of the locking magnetron still varied drastically when its load reflection coefficient deviated from 0.13 to 0.50. The decrement were 52.6% ($\rho = 0.075$), 76.2% ($\rho = 0.15$), and 72.3% ($\rho = 0.2$) in Status #2, respectively. Obviously, the variation tendency of the locking bandwidth is more similar to the prediction of the proposed model than the Adler's condition.

Fig. 7(b) depicts the effective external quality factor $Q_{\text{ext}}$ deduced by substituting the measured locking bandwidth, injected power, and output power into Adler's equation. The variation of the curves were similar to the estimation in Fig. 2, where the effective $Q_{\text{ext}}$ increased with load reflection level increased. Typically, the effective $Q_{\text{ext}}$ substantially increased in Status #1 with the injection ratios of 0.05 and 0.071, where effective $Q_{\text{ext}}$ increased from 150 to 1390, and from 106 to 3386, respectively. It is interesting to note that the effective $Q_{\text{ext}}$ of $|\Gamma| = 0.98$ in Status #1 with the injection ratio of 0.15 was smaller than other cases even if the locking bandwidth was narrow and output power was small. During the susceptance adjustment in a practical magnetron system, the extraneous modes would be excited to dissipate the energy in the coupled system (include magnetron and the load) [16], where the Adler's equation might not be suitable in the deduction of the effective $Q_{\text{ext}}$.

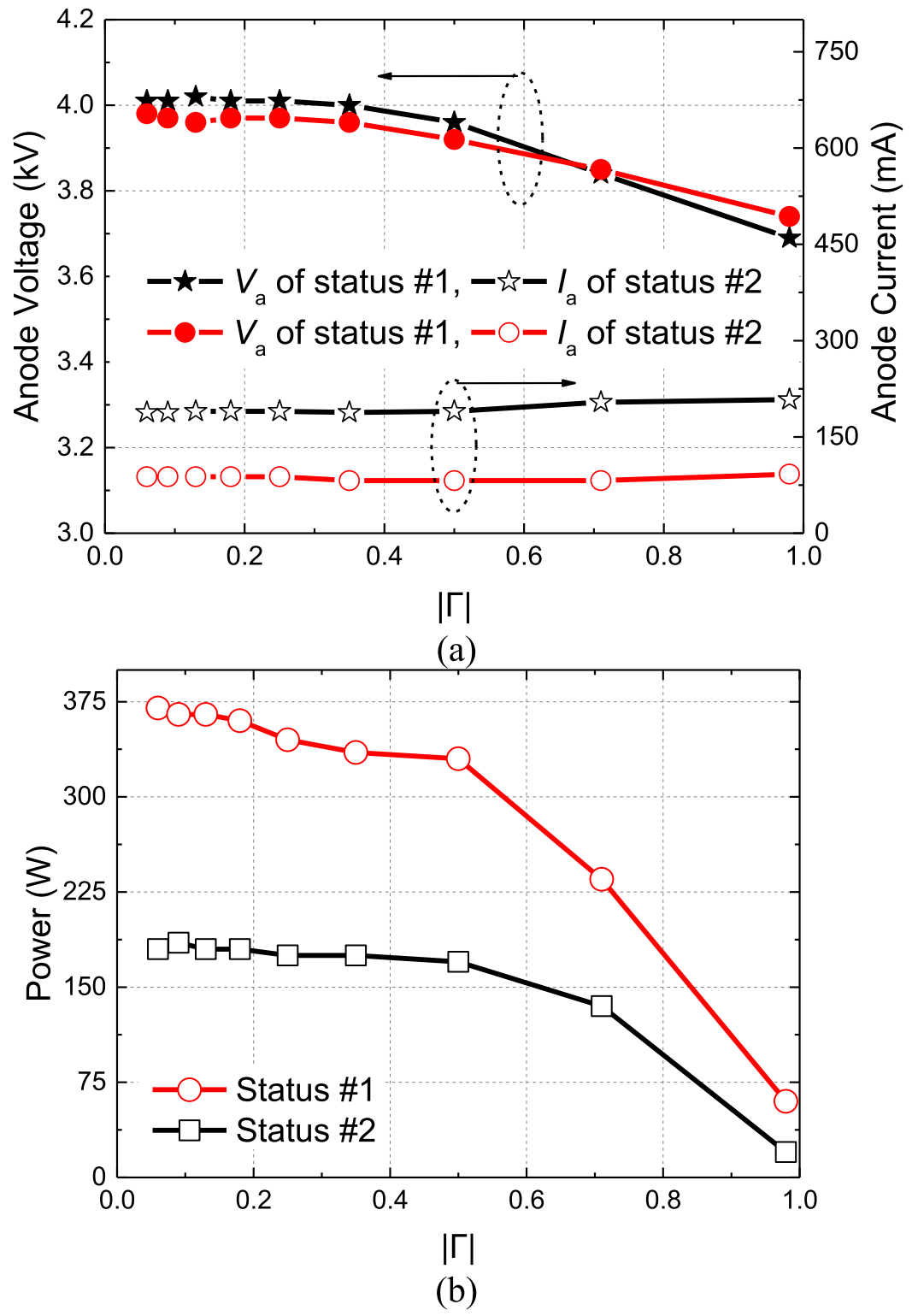

Fig. 8. Measured results of the operating parameters versus different load reflection levels. (a) Anode voltages and currents. (b) Power of both high- and low-power conditions.

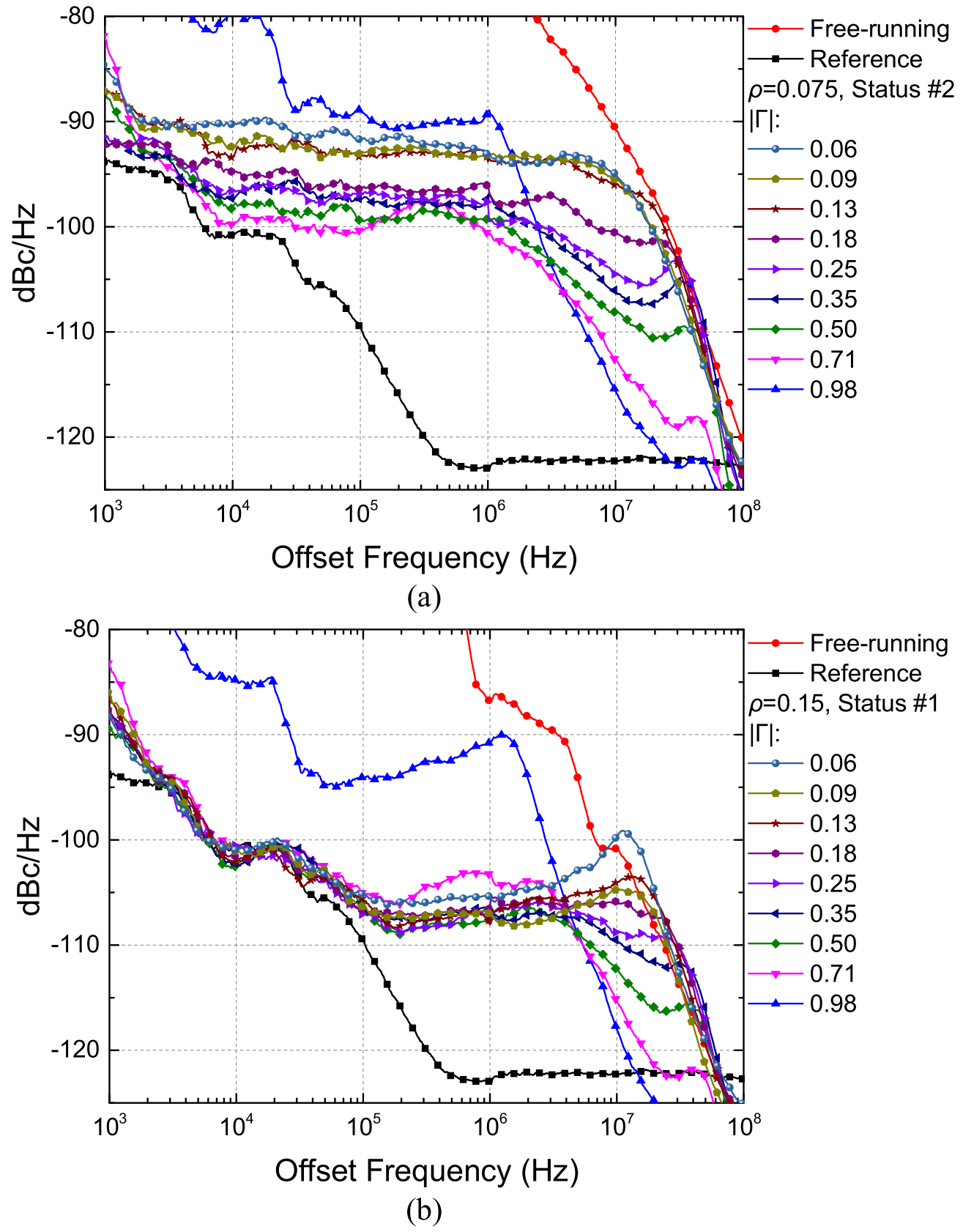

Fig. 9. Phase noise variations for the injection-locking magnetron with different load reflection levels. (a) $\rho = 0.075$. (b) $\rho = 0.15$.

Fig. 8 depicts the measured operating conditions of the magnetron (e.g., $V_a$, $I_a$, and $P_{\text{out}}$). The magnetron was driven using a current flow power supply whose output voltage will alter when the load property changes. Fig. 8(a) and (b) shows that with the increasing load reflection level, the driven current largely remained the same; however, the voltage gradually decreased. When $|\Gamma|$ varied from 0.35 to 0.98, the output power drastically decreased (Status #1: varied from 335 to 60 W; Status #2: varied from 175 to 20 W) even if the driven power was slightly varied (Status #1: varied from 752 to 768 W; Status #2: varied from 324 to 344 W). Whereas, when $|\Gamma|$ varied from 0.06 to 0.50, the dc-to-microwave conversion efficiency $\eta_{\text{dc}-\text{MW}}$ of the magnetron varied slightly. Varying from 49.1% to 43.9% in Status #1 and from 53.9% to 50.1% in Status #2, respectively. With $|\Gamma|$ further increasing, $\eta_{\text{dc}-\text{MW}}$ deteriorated drastically to the levels of lower than 10%. Both $\eta_{\text{dc}-\text{MW}}$ and deduced effective $Q_{\text{ext}}$ suggest that the coupled ability from the magnetron to the load was terribly weakened.

Fig. 9 shows the plots of the phase noise versus the offset frequency for the injection-locked magnetron with different load reflection levels. For the magnetron with injection ratios of 0.075–0.15 [see Fig. 9(a) and (b), respectively], the phase noise was reduced. In Fig. 9(a), the noise level dropped by the increasing reflection, where $|\Gamma|$ varied from 0.06 to 0.50. And the maximum noise suppression level was ~10 dB at the offset frequency band of 10 kHz to 1 MHz. Thereby, proper load reflection can significantly prohibit spurious output at the low-injection power condition. At low offset frequencies of less than 1 MHz, the suppression levels caused by the various load reflection levels were not easy to visually distinguish [see Fig. 9(b)]. However, phase noises deteriorated ~16 dB in Status #1 and ~10 dB in Status #2, when high reflection levels (e.g., $|\Gamma|$ varied from 0.50 to 0.98) were introduced; and curves of $|\Gamma| = 0.98$ also indicated that the magnetron almost failed to be locked. If the offset frequency was very high (i.e., exceeding 1 MHz), the phase noise of the injection-locked magnetron was suppressed by the load reflection, regardless of how the injection ratio changes. Herein, our measurements with respect to the offset frequency were qualitatively similar with the predictions in Section II.

Our investigated results and the consideration of the cost of actual applications show that the setup of the load reflection level should depend on the type of application. In the case of high-energy physics, a proper load reflection level is recommended instead of increasing the injection strength for higher output purity and precise phase control. The load reflection should be minimum to guarantee a high data rate in communication applications and improve the magnetron's output consistency.

## V. Conclusion

This study theoretically analyzed the effects of various load reflection levels on an injection-locked magnetron and experimentally verified them in a 5.8-GHz magnetron system. Various load reflection levels were introduced to the equivalent circuit model to ease the system behavior evaluation of the injection-locked magnetron. The magnetron performance was demonstrated when the load reflection was tuned using an adjustable-susceptance E–H tuner. The locking bandwidth of the magnetron was narrowed with the increasing load reflection coefficient. In contrast, a proper load reflection was accompanied by an improved noise performance. The measured results qualitatively agreed with the theoretical estimation.

Our investigations indicate that a tradeoff of the load reflection levels could be made in different advanced applications based on injection-locked magnetrons.